%% file: stopping.tex
\documentclass[%
reprint,
 amsmath,amssymb,
 pra,
]{revtex4-2}

\usepackage{graphicx}
\usepackage{dcolumn}
\usepackage{bm}
\usepackage[colorlinks=true,linkcolor=blue,citecolor=blue,urlcolor=blue]{hyperref}%
\usepackage[mathlines]{lineno}
\usepackage{comment} 
\usepackage{float}
\usepackage{amsmath,amssymb}
\usepackage[dvipsnames]{xcolor} 
\usepackage[normalem]{ulem}
\usepackage{MnSymbol,wasysym,ifsym}
\usepackage{soul}
\usepackage{xcolor}
\usepackage{multirow}
\usepackage{adjustbox}
\usepackage{booktabs}
\usepackage{siunitx}
\usepackage{float}

\begin{document}

\preprint{APS/123-QED}

\title{The $d$-electron contribution to the  stopping power of  transition metals}

\author{J. P. Peralta}
\email{jpperalta@iafe.uba.ar}
\affiliation{%
Instituto de Astronomía y Física del Espacio, CONICET and Universidad de Buenos Aires, Buenos Aires, Argentina.
}%

\author{A. M. P. Mendez}
\affiliation{%
Instituto de Astronomía y Física del Espacio, CONICET and Universidad de Buenos Aires, Buenos Aires, Argentina.
}%

\author{D. M. Mitnik}
\affiliation{%
Instituto de Astronomía y Física del Espacio, CONICET and Universidad de Buenos Aires, Buenos Aires, Argentina.
}%

\author{C. C. Montanari}
\affiliation{%
Instituto de Astronomía y Física del Espacio, CONICET and Universidad de Buenos Aires, Buenos Aires, Argentina.
}%

\date{\today}

\begin{abstract}
We present a new non-perturbative model to describe the stopping power by ionization of the $d$-electrons of transition metals. These metals are characterized by the filling of the $d$-subshell and the promotion of part of the electrons to the conduction band. The contribution of $d$-electrons at low-impact energies has been noted experimentally in the past as a break of the linear dependence of the stopping power with the ion velocity. In this contribution, we describe the response of these electrons considering the atomic \textit{inhomogeneous} momentum distribution. We focus on the transition metals of Groups 10 and 11 in the periodic table: Ni, Pd, Pt, Cu, Ag, and Au. Results are shown to describe the low 
energy-stopping power, with good agreement with the experimental data and with available TDDFT results.
By combining the present non-perturbative model for the $d$-subshell contribution with other approaches for the valence electrons and for the inner shells, we provide a coherent theoretical method capable of describing the stopping power of these transition metals from the very low to the high energy region.
\end{abstract}


\maketitle

\section{\label{sec:intro} Introduction}

Following IUPAC Principles of Chemical Nomenclature~\cite{IUPAC}, a transition metal is any element of the groups 3 to 12 of the periodic table, also known as the $d$-block elements.  Their electronic configuration is $ns^{k-j} \ +\ (n-1)d^j$, with $n$ being the main quantum number, $k$ the group number, and $j$ the number of $d$-electrons. In the literature (\cite{IUPAC} and references therein), there is a discussion if group 12 (Zn, Cd, and Hg) is a transition metal or not because of its filled $d$-subshell,  i.e. $s^2d^{10}$, and the related chemical characteristics.
For example, 
for the solids of group 12, the experimental $d$-binding energy has been measured (being $E_d \sim 10$ eV in all the cases~\cite{Williams:95}); however, for solids of groups 3 to 11, this value is still under discussion. The binding energy is an important quantity of the partially filled $d$ metals, which strongly impacts the electronic energy loss. 

For the \textit{earlier} transition metals ($k=3-7$), the maximum oxidation state is $k$~\cite{IUPAC}, i.e., all the $d$ and $s$ electrons may be valence electrons. In solid, all these $k$ electrons belong to the conduction band and are well-described by the free electron gas (FEG) model \cite{stopping_cr_mo_v_nb}. The Seitz radius $r_S$ of these elements also have experimental and theoretical values agreeing within $5\%$, which makes them \textit{canonical} metals~\cite{MONTANARI2017}.
For the \textit{later} transition metals ($k=8-11$), the oxidation number is much smaller than $k$~\cite{IUPAC}. As solid metals, the number of electrons in the FEG is not $k$ either: $d$-electrons are partially promoted to the FEG, and the rest remain bound. The number of $d$-electrons in the conduction band, the number of bound electrons, and the value of the ionization gap, $E_d$, are very relevant and still open subjects.

The energy loss in the \textit{later} transition metals has been the subject of multiple experimental \cite{Valdes2016,Markin2009,Cantero2009,Valdes1994,Markin2008,ROTH2018,Valdes1993,Go13,Cel2013,Cel15,Moller2002,BRUCKNER2018,VALDES2000,Pr2012} 
and theoretical studies~\cite{Denton2008,Abril2022,Zeb2012,Quashie2016,Quashie2018,Li2022,Li2023,Zhao2024}. These works were guided not only by their well-known properties and applications 
but also by some specific findings about low-energy stopping power and the response of $d$-electrons. At very low energies, the FEG approximations suggest a linear dependence of the electronic stopping power with the ion velocity.
For Cu, Ag and Au (group 11), the experimental values showed an unexpected break of this linear behaviour \cite{Valdes1993,Valdes1994,Cantero2009,Markin2008,Markin2009,Valdes2016,Go13,Cel2013,Cel15,ROTH2018}. This has been attributed to the contribution of $d$-electrons, as a drastic change in the $r_S$ above a specific critical impact velocity~\cite{Cantero2009}. 
Surprisingly, the low energy stopping power of Ni, Pd, and Pt (group 10) does not show the mentioned change in slope~\cite{Cel2013,Cel15,Moller2002,VALDES2000,BRUCKNER2018}. 

Perhaps the most detailed ab initio theory for stopping power calculations is the time-dependent density functional theory (TDDFT) \cite{Runge1984}. This theory accurately describes the low energy stopping power, being more suitable for channelling or monocrystalline targets than for polycrystalline or amorphous ones (off-channelling).
However, including all (or most of) the target electrons to extend the stopping power to intermediate and high-impact energies represents a 
heavy task. 
Recent TDDFT results for 
stopping power in Fe, Ni, Cu, Pt, and Au~\cite{Zeb2012,Quashie2016,Quashie2018,Li2022,Li2023,Zhao2024} show a soft non-linearity rather than the mentioned broken line and change in slope. 

In this work, we present a \textit{new} non-perturbative model for calculating the $d$-electron contribution to the electronic stopping power in the \textit{later} transition metals. The goal of this contribution is to describe the experimental values at low-impact energies but also the total electronic stopping power in an extended energy range. We focus on groups 10 and 11, Ni, Pd, Pt, Cu, Ag, and Au, due to the different low-energy behaviours experimentally found. We aim to understand and describe the physics involved in the response of these weakly bound $d$-electrons to the ion passage. 
The present results are compared with the experimental compilation in the IAEA Electronic Stopping Power Database~\cite{iaea,MONTANARI2024_IAEA}, highlighting the most recent low-energy experimental data \cite{Valdes1993,Valdes1994,Cantero2009,Markin2008,Markin2009,Go13,Cel2013,Cel15,ROTH2018,Moller2002,VALDES2000,BRUCKNER2018,Pr2012}. We also collate our curves with the available TDDFT (off-channelling) results~\cite{Quashie2016,Quashie2018,Li2022,Li2023}. 
By means of the independent shell approximation, 
we are able to predict the total electronic stopping power by combining the \textit{new} model results for the $d$-electron contribution, with those corresponding to non-perturbative~\cite{MONTANARI2017} and perturbative FEG models~\cite{mermin}, and the shellwise local approximation (SLPA) for the deeply bound shells~\cite{Montanari2013, Peralta2022}. The present results focus on the low energy region and the dependence of the stopping power with the impact velocity, and also cover the extended energy range up to $100$~MeV.

The new theoretical proposal is introduced in Sec.~\ref{sec:theory}. The present results and the comparison with the available experimental measurements are displayed and analysed in Sec.~\ref{sec:res}. Finally, the conclusions are given in Sec.~\ref{sec:concl}. Atomic units are used throughout this work unless explicitly mentioned.

\section{\label{sec:theory} Theoretical model}

Let us consider a point charge moving with velocity $v$ in a cloud of $d$-electrons with an \textit{inhomogeneous} velocity profile and a momentum distribution function $f(p)$. The electronic stopping cross-section, $S(v)$, is given by~\cite{NAGY1996,Wang-Nagy-Echenique-1998} 
\begin{equation} \label{eq:stopping_Nagy_1}  
S(v)=\frac{2}{(2 \pi)^3} \int \mathrm{d} \vec{p} \ f(p)\ v_r \frac{\vec{v}_r \cdot \vec{v}}{v} \sigma_{tr}\left(v_r\right),
\end{equation}
where $\vec{v}_r= \vec{v} - \vec{p}$ is the relative velocity, $\vec{p}$ is the electron momentum, which is equivalent to the electron velocity in atomic units, $\sigma_{tr}\left(v_r\right)$ is the transport cross-section
\begin{equation} \label{eq:str}
\sigma _{tr}(k)=\frac{4\pi }{k^{2}}\sum_{l=0}^{\infty}(l+1)\sin ^{2}%
\left[ \delta _{l}(k)-\delta _{l+1}(k)\right],   
\end{equation}%
and $\delta _{l}(k)$ are the phase shifts generated by a central potential $V(r)$ of the projectile screened by the $d$-electrons. In the present proposal, we consider the velocity-dependent screened potential introduced in Ref. \cite{MONTANARI2017}, which
verifies the cusp condition for the induced density around the charged projectile. 

By writing $v_r=\left( v^2\ + \ p^2-2\ v \ p \cos \ \varphi \right)^{1/2}$, invoking $d\vec{p}=2 \pi \ p^2 dp \sin \ \varphi \ d\varphi$, and changing the integration-variable from $\varphi$ to $v_r$, Eq.~(\ref{eq:stopping_Nagy_1}) can be expressed as 
\begin{equation} \label{eq:stopping_Nagy_4}
S(v) =\frac{1}{(2 \pi v)^2} \int_0^{\infty} dp \ p \ f(p) \ I(p),
\end{equation}
with
\begin{equation} \label{eq:Ip}
I(p)=\int_{\left|v-p\right|}^{\left|v+p\right|} d v_r v_r^2\left(v_r^2+v^2-p^2\right) \ \sigma_{tr}\left(v_r\right).
\end{equation}
A change of the order of the integrals in Eqs.~(\ref{eq:stopping_Nagy_4}) to (\ref{eq:Ip}) gives
\begin{equation} \label{eq:nagy4_new}
S(v) =\frac{1}{(2 \pi v)^2} \int_0^{\infty} d v_r \ v_r^2 \ \sigma_{tr}\left(v_r\right) \ I'(v_r),
\end{equation}
with
\begin{equation} \label{eq:Ivr}
I'(v_r)=\int_{\left|v_r-v\right|}^{\left|v_r+v\right|} \left(v_r^2+v^2-p^2\right) \ p \ f(p)\ d p.
\end{equation}
This change locates the distribution function in the inner integral and the transport cross-section in the outer one, improving the performance of the numerical calculations of the stopping cross-sections.

In the present model, we propose to describe the response of $d$-electrons through an \textit{inhomogeneous} momentum distribution function given by
\begin{equation}\label{eq:trf}
    f_{nl}(p)= \frac{(2 \pi)^3}{2}\ |\Phi_{nl}(\vec{p})|^2, 
\end{equation}
with $\Phi_{nl}(\vec{p})$ being the Fourier transform of the 
wave functions $\phi_{nl}(\vec{r})$, normalized to the number of electrons $N_{e}$ in the $nl$-subshell.
With this definition, the distribution function $f_{nl}(p)$, verifies
\begin{equation} \label{eq:fp}
    \int f_{nl}(p) \ d\vec{p}=\frac{(2 \pi)^3}{2} N_{e}.
\end{equation}

In the \textit{homogeneous} FEG model~\cite{NAGY1996}, the momentum distribution function is given by a step function within the Fermi sphere, $f_{\mathrm{FEG}}(p)= \Theta(p-p_F)$,  where $p_F$ is the Fermi momentum, and
\begin{equation} \label{eq:theta}
\int \Theta(p-p_F) \ d\vec{p}=\frac{(2 \pi)^3}{2} n_e,
\end{equation}
with  $n_{e}$ being the density of electrons, $n_{e}=N_{\mathrm{FEG}} \ n_{at}$, $N_{\mathrm{FEG}}$ is the number of electron in the FEG,  and $n_{at}$ is the atomic density. The difference between Eqs. (\ref{eq:fp}) and (\ref{eq:theta}) is the difference between stopping cross section and energy loss per unit path length.

Within this \textit{new} model, given by Eqs. (\ref{eq:nagy4_new}) to (\ref{eq:trf}), we analyze the contribution to the stopping power of the $3d$, $4d$, and $5d$ subshells of Ni, Cu, Pd, Ag, Pt, and Au, as applicable. The wave functions employed for Ni and Cu are the Hartree-Fock ones~\cite{BUNGE1993}, while for Pd, Ag, Pt, and Au, we solved the fully relativistic atomic structure using the {\sc hullac} code package~\cite{Klapisch1977,Bar-Shalom2001} (for more details, see Refs.~\cite{MENDEZ_Relativistic}).
The momentum distribution functions $f_{nl}(p)$  given by  Eq. (\ref{eq:trf}), 
were obtained analytically by expanding the wave functions $\phi_{nl}(r)$ in Slaters. By considering this expansion, the Fourier transform of the Slater functions is analytical using Flannery-Levy integrals~\cite{Flannery}.  Then, the integration in Eq.~(\ref{eq:Ivr}) is also analytical (see Appendix \ref{App} for details).

\section{\label{sec:res} Results}

\begin{table*}[t]
\caption{\label{table_period} Atomic structure, valence, and $d$-electrons of Ni, Cu, Pd, Ag, Pt, and Au: The outer electrons of the electronic configuration are shared, $N_{\mathrm{FEG}}$ electrons are in the FEG, and $N_{d}$ remain in the $d$-subshell; $r_S$, is the Wigner-Seitz radii; $\omega_p$, the plasmon frequency,  $E_F$, the Fermi energy. It also included the experimental values for the plasmon peak, $\omega_p^{\mathrm{exp}}$, and the dump value, $\gamma_p^{\mathrm{exp}}$ obtained from \cite{ELF_Werner}. $E_d$,  $v_d$ and $<r_d>$ are the theoretical binding energy, mean velocity, and mean radio of the atomic $d$-subshell from Hartree-Fock \cite{BUNGE1993} (Ni and Cu) and full relativistic calculations \cite{MENDEZ_Relativistic} (Pd, Pt, Ag, Au). Atomic units are used. 
}.
\begin{ruledtabular}
\begin{tabular}{lccl|cccccc|cccc} & Element
            & $Z$  & Atomic   &   &  & FEG & & & &  & & $d$-bound & \\
           & &   & configuration \    & $N_{\mathrm{FEG}}$    & $r_S$  & $\omega_p$ & $\omega_p^{\mathrm{exp}}$ & $\gamma_p^{\mathrm{exp}}$  & $E_F$ \   & $N_d$ &   $E_d$ &  $v_d$ & $<r_d>$\\
\hline
\quad Group 10 & Ni & 28 & [Ar]~$3d^8\,4s^2$         \qquad& 3 & 1.80 & 0.714 & 0.716 & 0.27 & 0.564 & 7 & 0.707 & 3.72&0.965\\
\quad & Pd & 46 & [Kr]~$4d^{10}$       \qquad& 7 & 1.50 & 0.942 & 0.937& 0.21& 0.815 & 3 & 0.216 & 2.60 &1.61\\
\quad & Pt & 78 & [Xe]~$5d^9 \,6s^1$      \qquad& 7 & 1.51 & 0.929   & 0.919 & 0.17 & 0.801 & 3 & 0.250 & 2.45& 1.71 \\
\quad Group 11 & Cu & 29 & [Ar]~$3d^{10}\,4s^1$ \qquad & 3 & 1.85 & 0.689 & 0.707 & 0.29 & 0.537 & 8 & 0.491 & 3.73 & 0.991\\
\quad & Ag & 47 & [Kr]~$4d^{10}\,5s^1$  \qquad& 3 & 2.09 & 0.572 &0.625 &0.20 & 0.420 & 8 & 0.641 & 2.88&1.37\\
\quad & Au & 79 & [Xe]~$5d^{10}\,6s^1$     \qquad& 7 & 1.57 & 0.877 & 0.864& 0.53 & 0.742 & 4 & 0.309 & 2.67 & 1.58
\end{tabular}
\end{ruledtabular}
\end{table*} 

In Table~\ref{table_period}, we tabulated the details about the transition metals studied here. These metals have loosely bound $d$-electrons. Some of them are promoted to the conduction band, while others remain bound to the target nucleus. Knowing how many $d$-electrons are part of the FEG and how many remain bound is crucial to the energy loss description. To this end, we analyzed data from reflection electron energy-loss spectroscopy in solids by Werner et al.~\cite{ELF_Werner}. The experimental plasmon frequency $\omega_p^{\mathrm{exp}}$ and dump $\gamma_p^{\mathrm{exp}}$ in Table~\ref{table_period} were obtained from the first significant peak and width of the energy loss function in Ref.~\cite{ELF_Werner}. From $\omega_p^{\mathrm{exp}}$, the number of electrons in the FEG can be inferred. We define $N_{\mathrm{FEG}}$ as the integer number that is closest to that value. The plasmon frequency $\omega_p$, the Wigner-Seitz radii $r_S$, and the Fermi energy $E_F$ in Table \ref{table_period} are obtained from these $N_{\mathrm{FEG}}$ values.

The atomic configuration and characteristics of the $d$-bound electrons in the columns on the right of Table \ref{table_period} are the result of the Hartree-Fock (Ni, Cu) and the fully relativistic atomic structure calculations (Pd, Ag, Pt, and Au)~\cite{Klapisch1977,Bar-Shalom2001,MENDEZ_Relativistic}. 
It is important to note that these values correspond to the atoms, not to the solid targets. This distinction is evident when comparing the values of $E_F$ and $E_d$ (the atomic binding energy of the $d$ electrons) displayed in Table~\ref{table_period}. Assuming that part of the $d$-electrons are promoted to the FEG, the remaining ones must have $E_d>E_F$. 
According to the values shown in Table \ref{table_period},  $E_d<<E_F$ for Pd, Pt, and Au, which suggests that for these targets, most of the loosely bound $d$-electrons are transferred to the FEG. Instead, in Ni, Cu, and Ag, only one or two $d$-electrons are promoted to the conduction band.
For solid Cu, Ag, and Au, density of states (DOS) calculations with DFT~\cite{Lin2008} are such that $0.05 \leq E_d-E_F \leq 0.15$. For Ni and Pt, the energy gap between the $d$ electrons and the Fermi energy is even smaller.

\begin{figure*}
\includegraphics[width=0.85\textwidth]{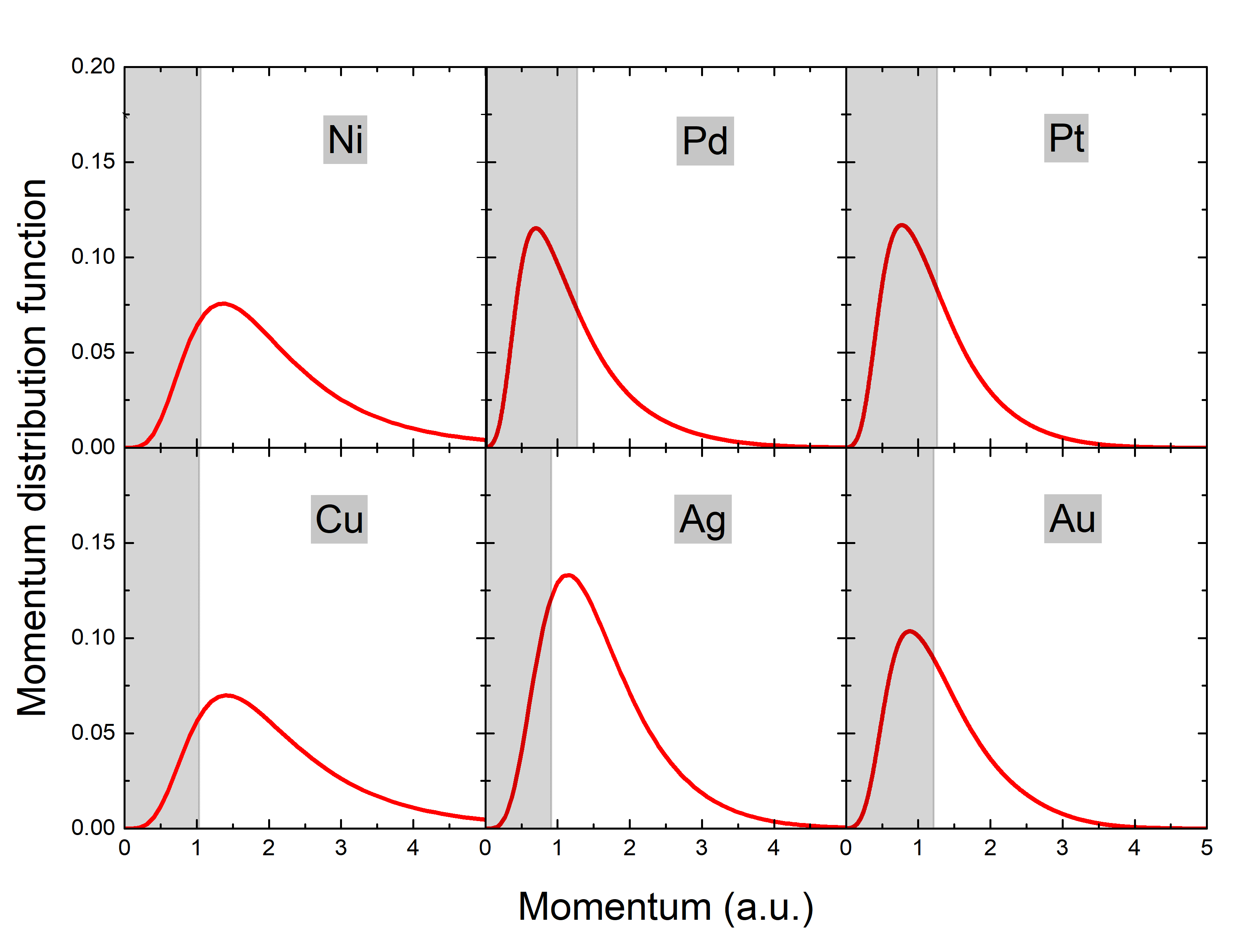}
\caption{
(Color online) Distribution functions of the $d$-electrons, 
$f_{nd}(p)$, for the six targets considered here (red-solid curve), and the FEG (light-grey area from 0 to Fermi momentum). 
}
\label{densAg} 
\end{figure*}

In Fig.~(\ref{densAg}), we display the \textit{inhomogeneous} momentum distribution functions $f_{nd}(p)$ given by Eq. (\ref{eq:trf}) for the six targets studied here. We also include in this figure the \textit{homogeneous} FEG distribution $f_{\mathrm{FEG}}(p)$. To compare them equivalently, we plotted $f_{nd}(p)\ n_{at}$. The Heaviside function in $f_{\mathrm{FEG}}$ is constant and equal to $1$, cutting off at $p_F$. Conversely, the $f_{nd}(p)$ distribution has a lower amplitude than the FEG one but extends far away from the Fermi sphere of the FEG and falls drastically within the Fermi region. For Pd, Pt and Au, the value of $v_d$ is smaller than for the others, which explains the maximum of their $f_{nd}(p)$ being shifted to lower values of $p$.

In Section~\ref{subsec:low-v}, we present our low-energy results for Ni, Cu, Pd, Ag, Pt, and Au and, in Section~\ref{subsec:extended}, we extend the calculation to high energy region. 
For proton impact energies around the maximum and below, the experimental stopping power cross sections present considerable dispersion~\cite{MONTANARI2024_IAEA}, and this spread is more notorious for the historically most measured targets (Au, Ag, Cu, Ni). We consider all the experimental data available in Ref.~\cite{iaea}; however, we pay special attention to the most recent measurements at low energies, namely, those conducted from 1990 onwards with $v\leq 1$. These values are highlighted in Figs. \ref{hniv} to \ref{hau_E} with colours and filled symbols. The other low-energy experimental data, which were measured 35 or more years ago, are displayed altogether with half-filled circles, while the rest of the data in the database~\cite{iaea}, i.e., the experiments at intermediate and high energies, are illustrated with empty circles.

\subsection{Low energy stopping cross sections}
\label{subsec:low-v}

\begin{figure}[h]
\includegraphics[width=0.45\textwidth 
]{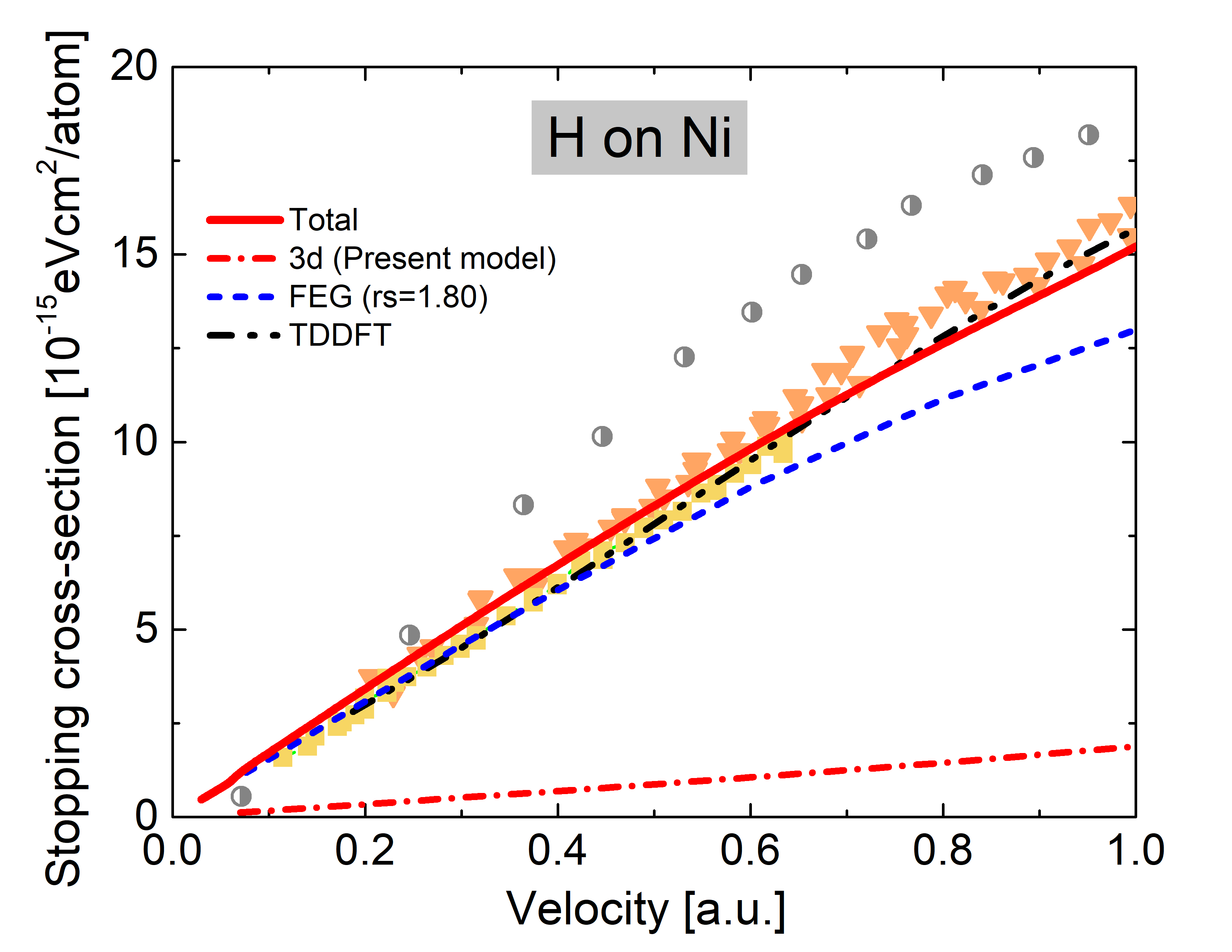}
\caption{\label{hniv} 
(Color online) Low-energy stopping cross-section of Ni for H as a function of the impact velocity. 
Curves: total stopping (red-solid line), present model for $d$-electron contributions (red dash-dotted line), the FEG stopping (blue-dashed line), and TDDFT results by Quashie \textit{et al.}~\cite{Quashie2018} (dark-dash-double-dotted line).
Symbols: 
$\filledmedtriangledown$ \cite{Moller2002}, $\blacksquare$ \cite{BRUCKNER2018}, {$\RIGHTcircle$} low energy data in \cite{iaea} previous to 1990. 
}
\end{figure}

\begin{figure}[h]
\includegraphics[width=0.45\textwidth
]{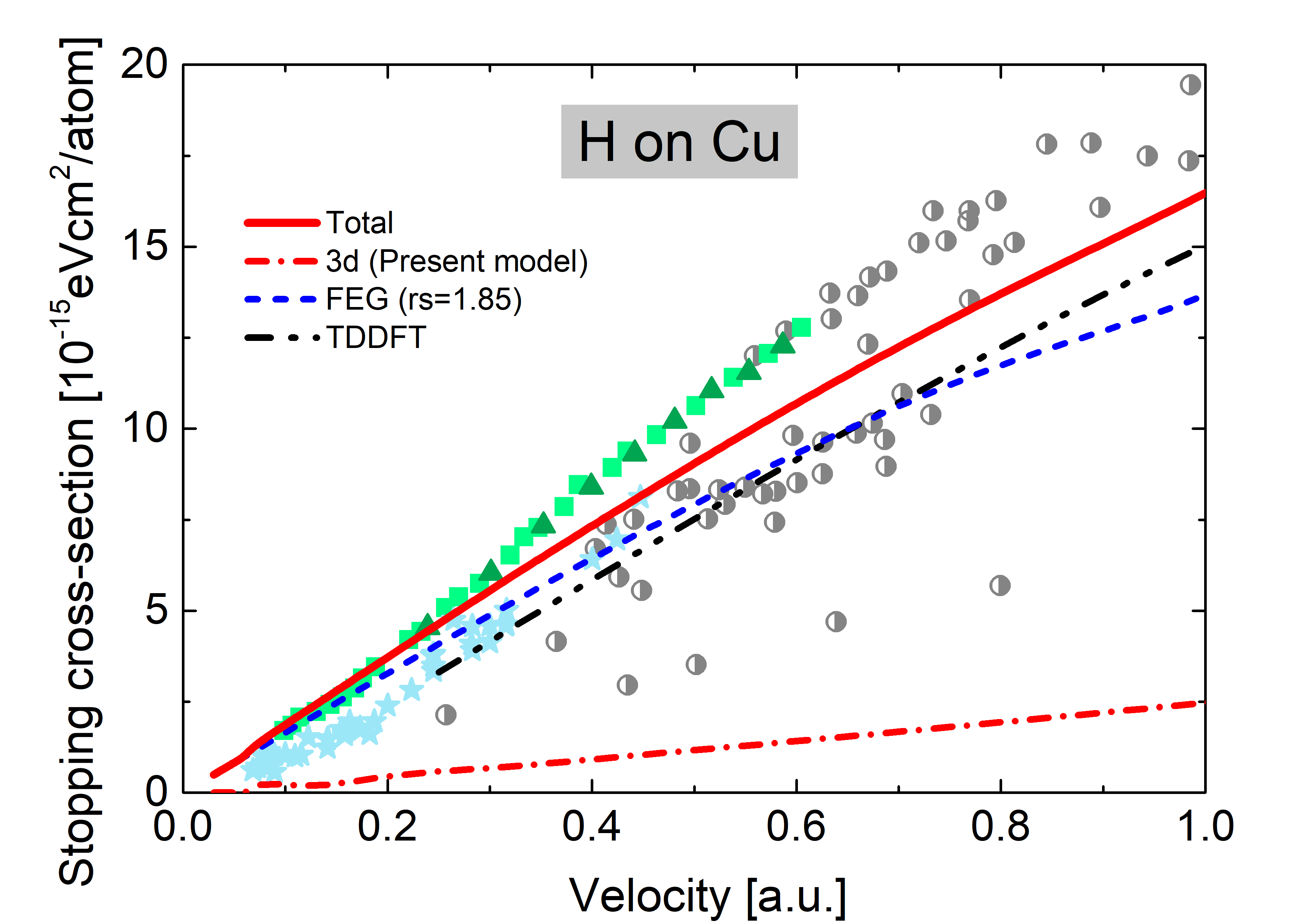}
\caption{\label{hcuv} 
(Color online) Low-energy stopping cross-section of Cu for H as a function of the impact velocity. 
Curves: as in Fig.~\ref{hniv}, TDDFT results by Quashie~\cite{Quashie2016}.
Symbols: $\bigstar$ \cite{Markin2009}, $\filledmedsquare$ \cite{Cantero2009}, $\filledmedtriangleup$ \cite{Valdes1994}, 
{$\RIGHTcircle$} data in \cite{iaea} previous to 1990.
}
\end{figure}

\begin{figure}[h]
\includegraphics[width=0.45\textwidth
]{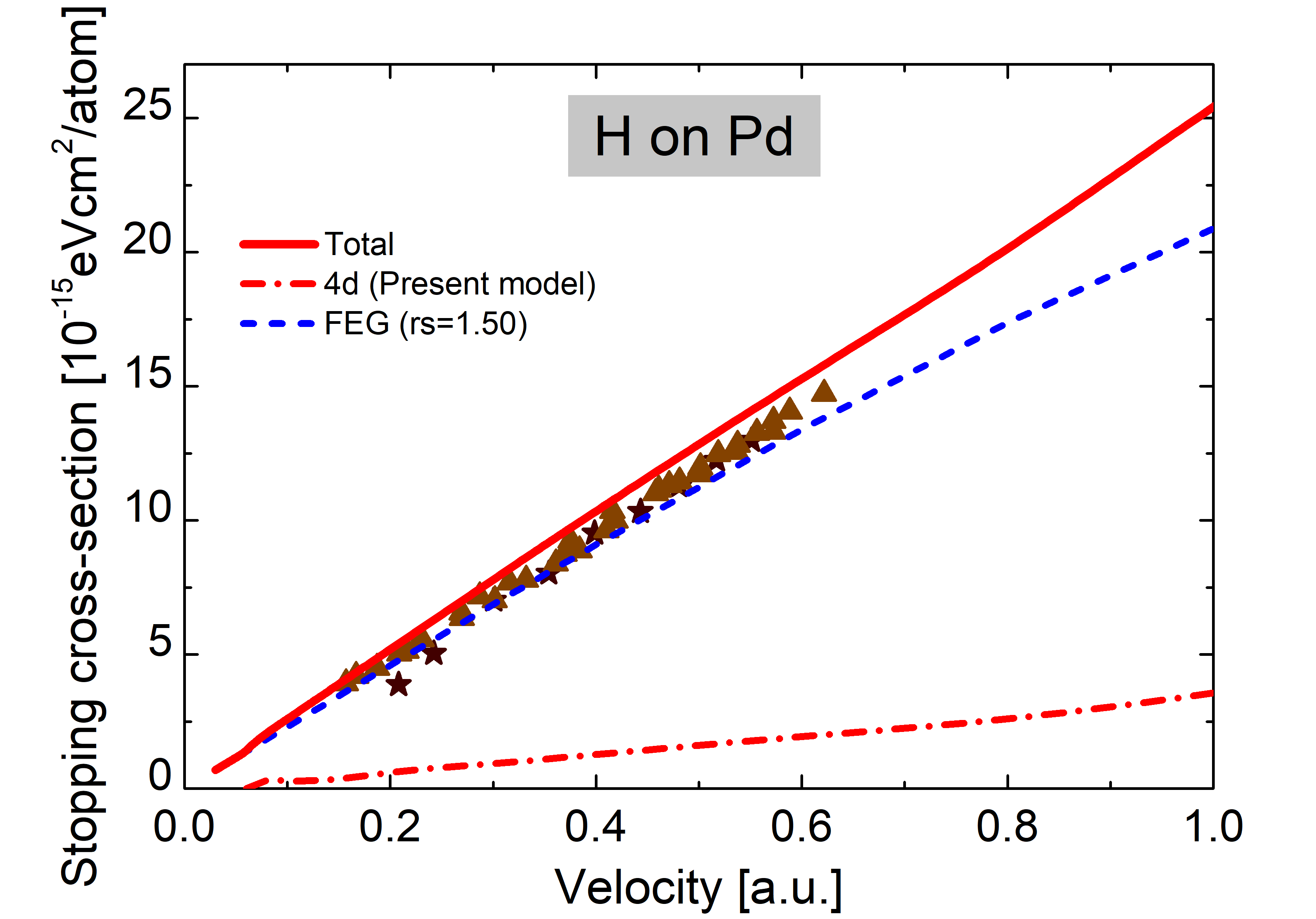}
\caption{\label{hpdv} 
(Color online) Low-energy stopping cross-section of Pd for H as a function of the impact velocity. 
Curves: as in Fig.~\ref{hniv}
Symbols: $\filledmedtriangleup$  \cite{Cel2013}, $\bigstar$ \cite{VALDES2000}.
}
\end{figure}

\begin{figure}[h]
\includegraphics[width=0.45\textwidth]{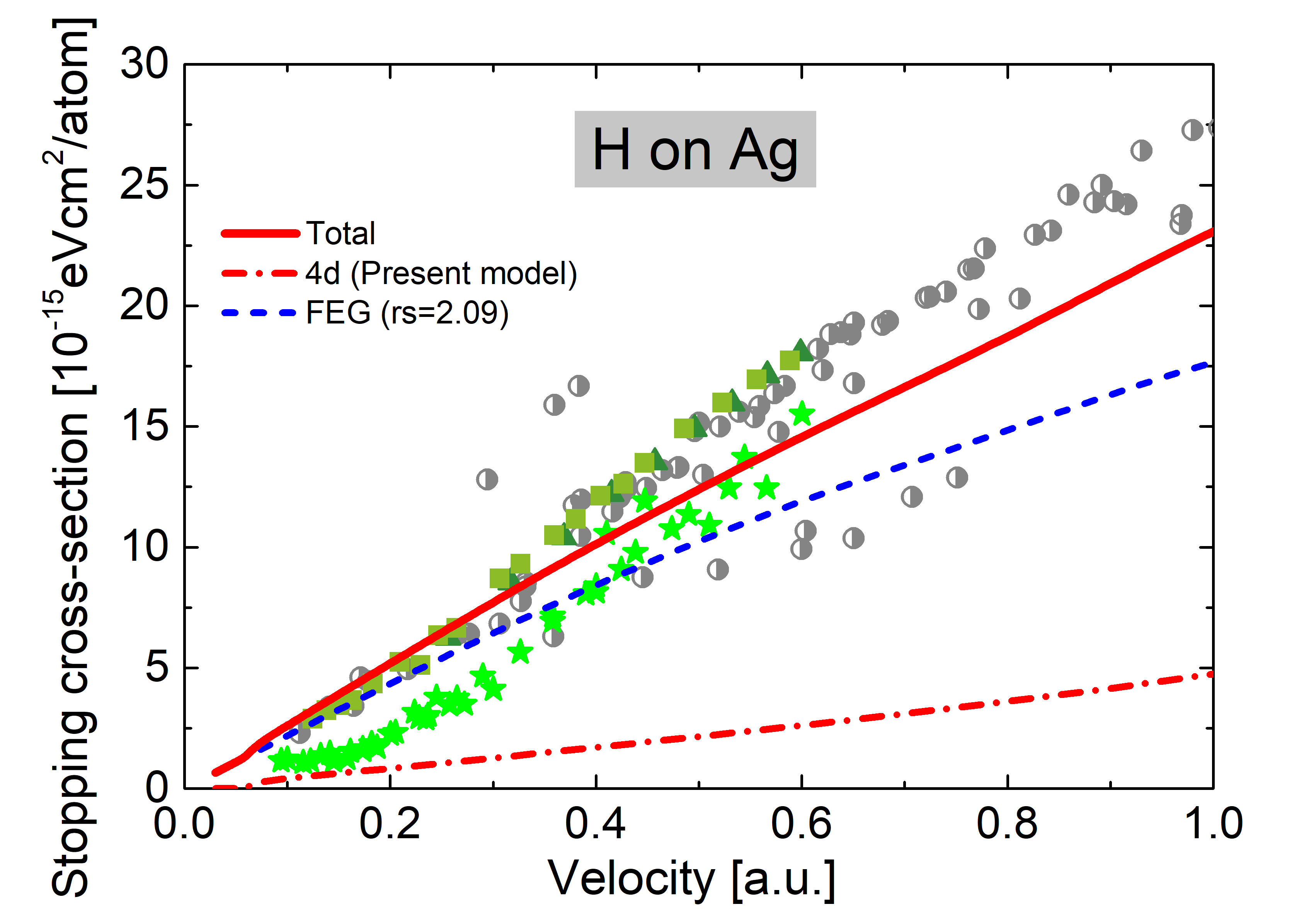}
\caption{\label{hagv} 
(Color online) Low-energy stopping cross-section of Ag for H as a function of the impact velocity. 
Curves: as in Fig.~\ref{hniv}.
Symbols: $\blacksquare$ \cite{Cantero2009}, $\bigstar$ \cite{Go13}, $\filledmedtriangleup$ \cite{Valdes1993},
{$\RIGHTcircle$} data in \cite{iaea} previous to 1990. }
\end{figure}

\begin{figure}[h]
\includegraphics[width=0.45\textwidth]{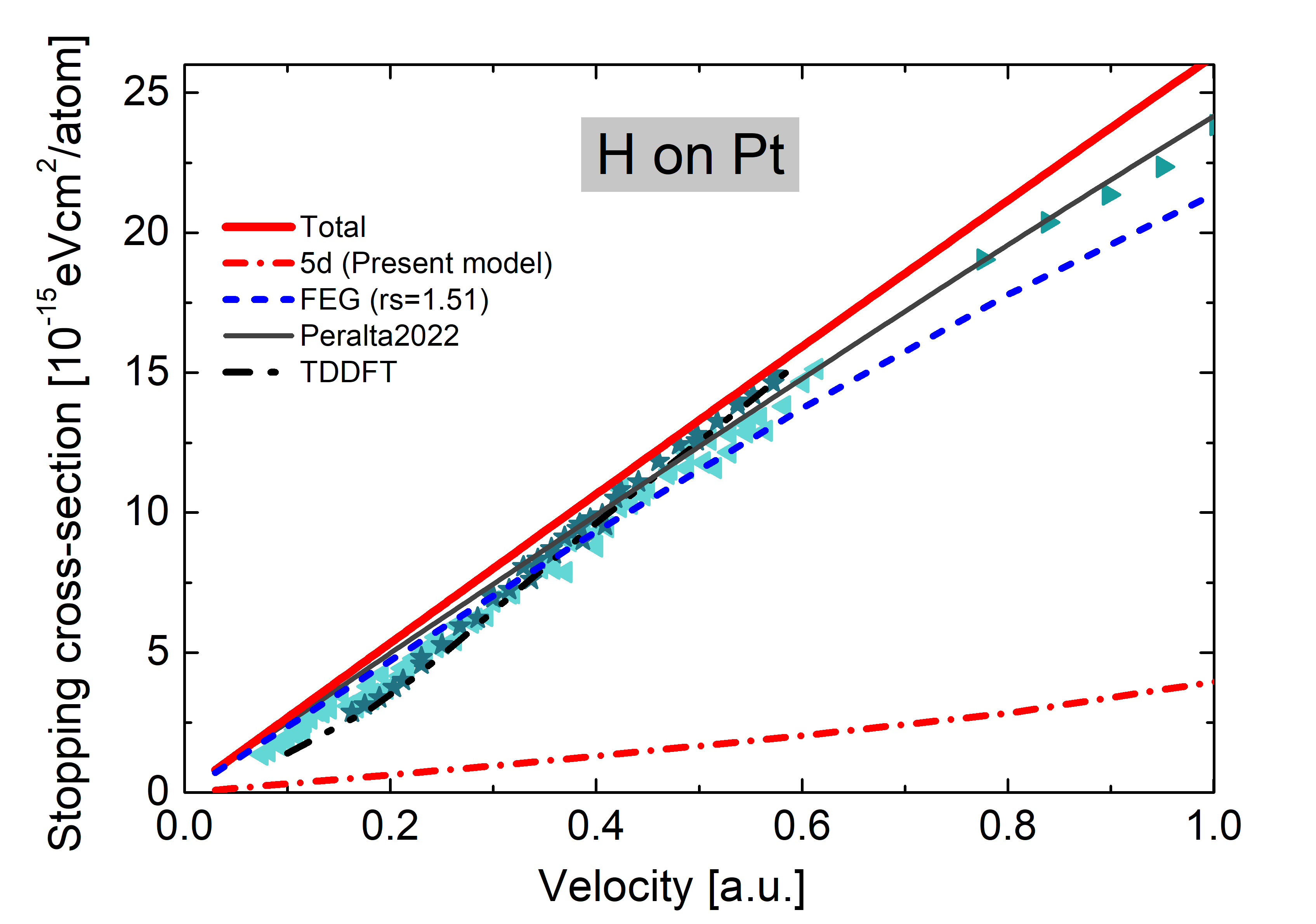}
\caption{\label{hptv} 
(Color online) Low-energy stopping cross-section of Pt for H as a function of the impact velocity. 
Curves: as in Fig. \ref{hniv}, with TDDFT results by Li\cite{Li2023}; 
dark grey-solid line, total values with all the $d$-electrons in the FEG by Peralta (2022) \cite{Peralta2022}.
Symbols: $\filledmedtriangleleft$ \cite{Go13}, $\bigstar$ \cite{Cel15} and $\filledmedtriangleright$ \cite{Pr2012}.
}
\end{figure}

\begin{figure}[h]
\includegraphics[width=0.45\textwidth
]{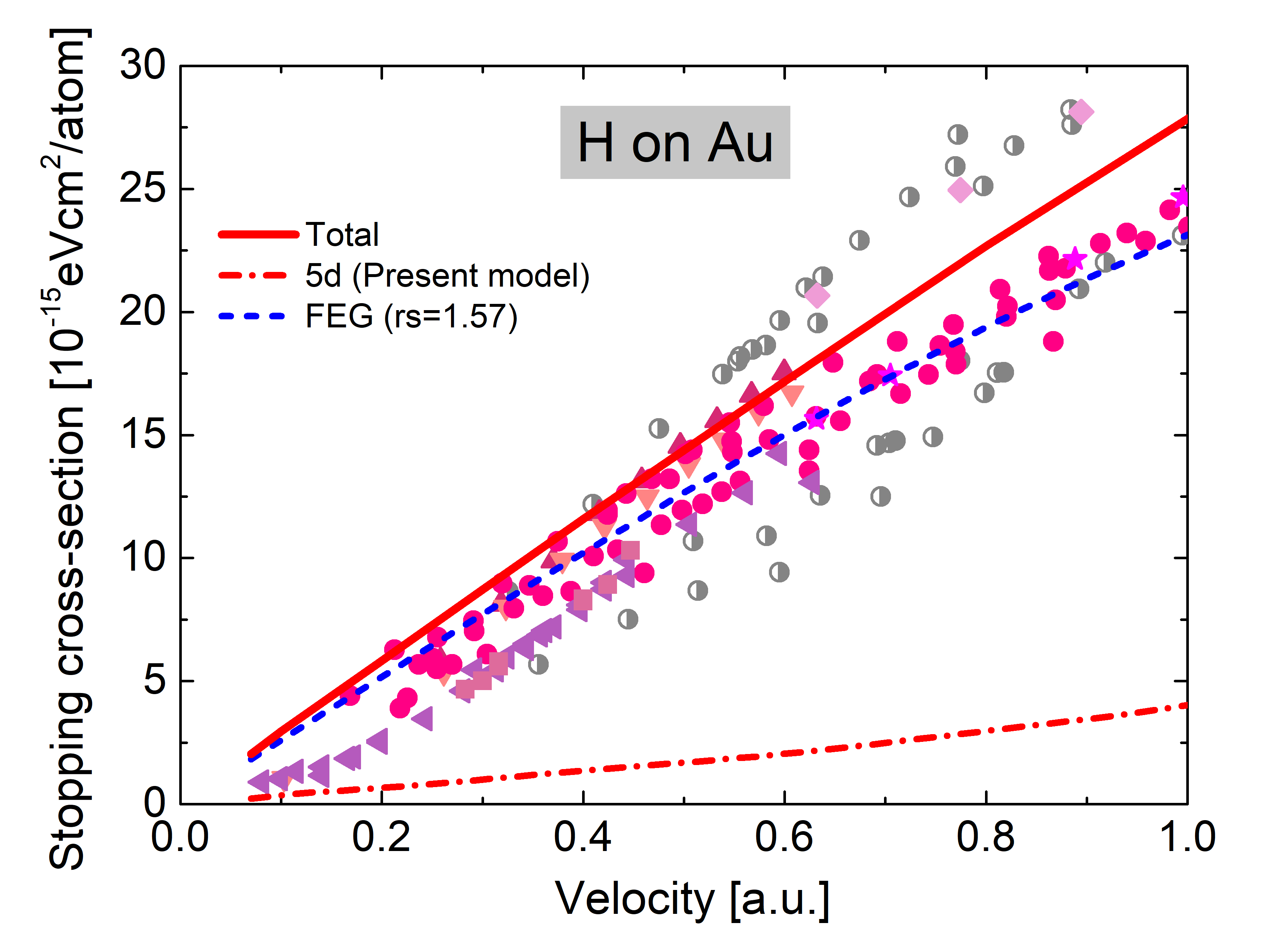}
\caption{\label{hauv} 
(Color online) Low-energy stopping cross-section of Au for H as a function of the impact velocity. 
Curves: as in Fig. \ref{hpdv}.
Symbols: $\CIRCLE$  \cite{Moller2002}, $\blacksquare$  \cite{Markin2009},  
$\filledmedtriangledown$  \cite{VALDES2000}$\filledmedtriangleup$ \cite{Valdes1993}, $\bigstar$ \cite{Pr2012}, $\filledmedtriangleleft$~\cite{Markin2008}, 
 $\blacklozenge$~\cite{MT1996},
{$\RIGHTcircle$} data in \cite{iaea} previous to 1990.
}
\end{figure}

In Figs. \ref{hniv} to \ref{hauv}, we display the present theoretical stopping cross sections of the six targets studied here as a function of the impact velocity $v$, with $0\leq v\leq1$, and we compare them with the available experimental data compiled in ~\cite{iaea}. The FEG and $d$-electron contributions are shown separately, and the total stopping cross-section results from adding both. 
The influence of the $d$-electron contribution to the stopping power is evident in all these figures. 

The stopping power cross sections of H in Ni are shown in Fig. \ref{hniv}. The importance of the $d$-contribution is remarkable above $v=0.5$. The present results allow describing the experimental values in this energy region. The agreement with data in Refs.~\cite{Moller2002,BRUCKNER2018} is very good. 
We also included in Fig. \ref{hniv} the TDDFT results by Quashie \textit{et al.}~\cite{Quashie2018} for off-channelling Ni, with 10 active electrons. 
The agreement between the present results and the TDDFT ones is excellent. We consider that the comparison of our results with the experimental 
data and an \textit{state of the art} model supports the present model. 

In Fig.~\ref{hcuv}, we illustrate the case of H in Cu. The spread of experimental data in this region is significant. At very low impact velocities, there are two tendencies of the experimental values, where the backscattering measurements from Ref.~\cite{Markin2009} are much lower than the transmission measurements from Refs.~\cite{Cantero2009,Valdes1994}. It can be noted that these two groups of data are supported separately by older measurements, \cite{No1975} and \cite{Mo1968}, respectively. Our total stopping cross sections, which adds the FEG 
and the $3d$ contribution, agree with Ref.~\cite{Cantero2009} at the lowest impact velocities and lays between both experimental data groups for $v>0.3$. As in the previous figure, we included the TDDFT values by Quashie \textit{et al.}~\cite{Quashie2016}. Our results are close but above these values, which 
only agree with the measurements from Ref. \cite{Markin2009} for $ v<0.3$. 
Our model for the $d$-electron contribution to the stopping power does not include an explicit energy gap; it solely relies on the information of the \textit{inhomogeneous} momentum distribution function. This missing feature may introduce
some degree of overestimation at very low velocities.

The present results for H in Pd are displayed in Fig. \ref{hpdv}. Only two sets of data are available at low velocities, \cite{Cel2013} and \cite{VALDES2000}, which agree pretty well among them and show an almost perfect linear dependency of the stopping power with the impact velocity. Our results describe these values quite well; the total stopping power is slightly larger than the experimental data, with differences that are less than $5\%$.

In Fig. \ref{hagv}, we display the case of H in Ag. Similarly to the Cu case, the experimental data is separated into two groups: the lower-lying data are backscattering measurements by Goebl \textit{et al.} \cite{Go13}, and upper-lying are transmission measurements by Cantero \textit{et al.} \cite{Cantero2009} and Valdes \textit{et al.} \cite{Valdes1993}. 
Present total stopping results are closer to the upper experimental values from Ref.~\cite{Cantero2009}
for $v<0.4$, and to the data from Ref.~\cite{Go13}
for $0.4 < v\leq 0.6$. Our results show the importance of the $d$-subshell contribution above $v=0.3$. As in the other cases, the present model does not describe the inclusion of $d$-electrons to the total stopping power as a sharp change of slope in the linear velocity dependence but as a smooth difference between the FEG contribution and the total stopping values. 

Present results for H in Pt are displayed in Fig. \ref{hptv}, together with the available data~\cite{Go13,Cel15,Pr2012} and recent TDDFT values by Li \textit{et al.}~\cite{Li2023,Li2022}. Our total values, considering the FEG and the $5d$ contributions, are slightly above these state-of-the-art values, but they are very close in amplitude. It is worth noting that the two experimental data sets for very low velocities \cite{Go13,Cel15} agree pretty well between them, and none of them indicates a break of the linear response at low energies. We also show in Fig.~\ref{hptv} our previous theoretical results, where we considered $N_{\mathrm{FEG}}=10$, i.e., all $d$-electrons are included in the FEG~\cite{Peralta2022}.  The difference is not large, but it shows a better overall agreement with the available experimental data. Nevertheless, we consider the present model to describe this group of targets more properly and in a more general way.

Finally, our results for H in Au are displayed in Fig.~\ref{hauv}. This system features the most significant number of stopping power measurements with the widest dispersion~\cite{iaea,MONTANARI2024_IAEA}. 
The present results are in good agreement with one group of data sets~\cite{VALDES2000,Valdes1993}, while they overestimate other groups of measurements~\cite{Markin2008,Markin2009}. 
The low energy measurements in these last measurements~\cite{Markin2008,Markin2009} are very close to channelling stopping values in Au $\langle 100\rangle$ by Valdes \textit{et al.}~\cite{Valdes2016}  and to the TDDFT results in Au $\langle 100\rangle$ by Zeb and collaborators~\cite{Zeb2012}. TDDFT calculations for polycrystalline Au could bring some clarity to this case. 

Considering the results in Figs. \ref{hniv} to \ref{hauv}, we can say that there is a clear difference between the transition metals of groups 10 and 11. The change in the linear behavior with the impact velocity is not necessarily the most important difference found in the cases examined here but rather the dispersion of experimental data, which is very important in group 11 and almost negligible in group 10. Our results agree with the measurements in metals from group 10, and somewhere in the middle of the data cloud of the metals from group 11. As already mentioned, we do not rule out a small overestimation at the lowest velocities shown here related to the band gap. This topic is the subject of future developments.

We thoroughly analyzed the values of $N_{\mathrm{FEG}}$ and $N_d$ in the six cases studied here; nevertheless, we acknowledge they may be the subject of discussion. To further examine the present model along with the choice of these electron numbers, we extend the present results to intermediate to high energies and compare them with all the available data.


\subsection{Stopping cross sections in an extended energy range}
\label{subsec:extended}

\begin{figure}[h]
\includegraphics[width=0.45\textwidth 
]{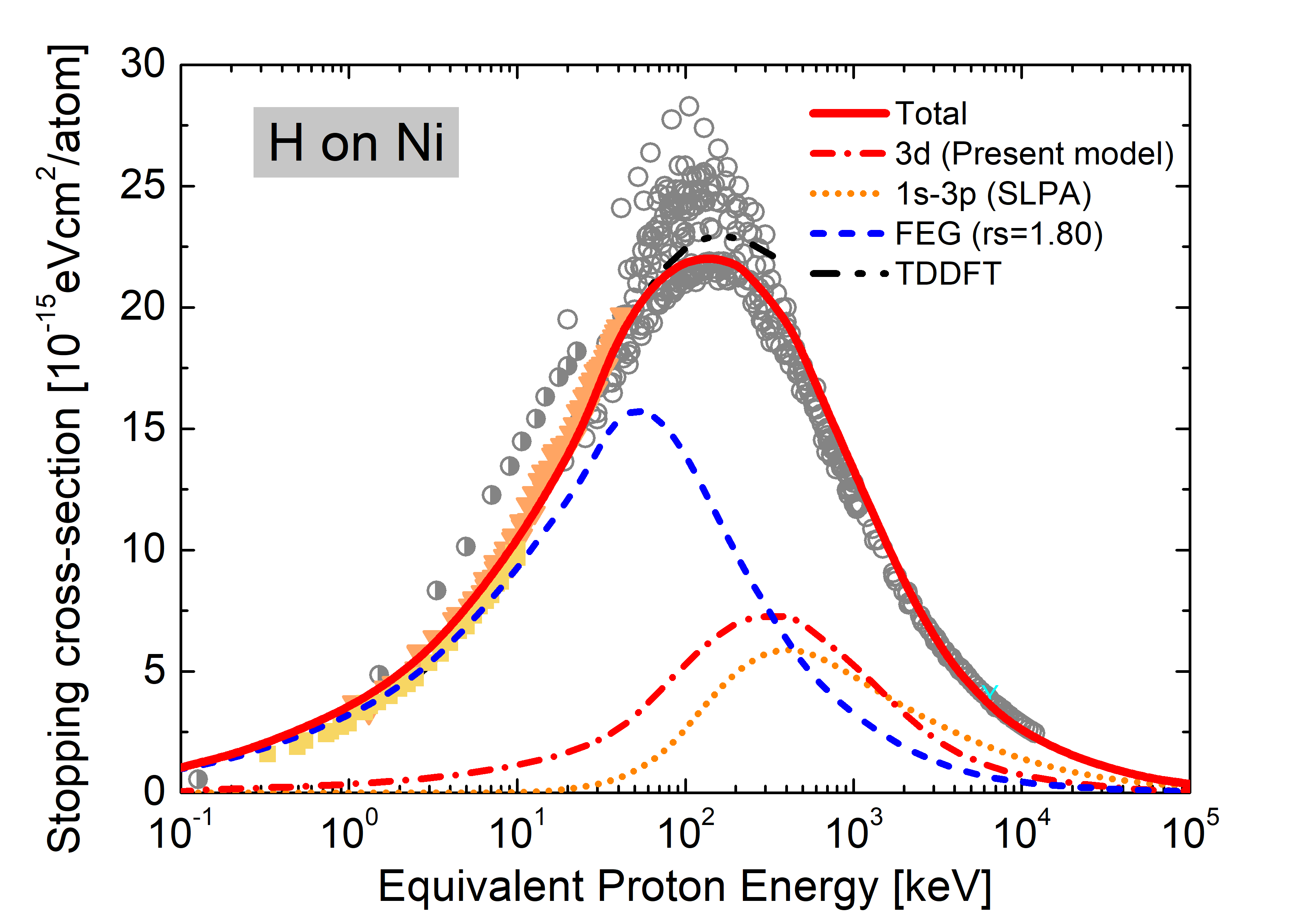}
\caption{\label{hni_E} Stopping cross-section of Ni for H as a function of the impact energy.
Curves: red-solid line, total stopping; red-dashed-dot, present model for $d$-electron contributions, blue-dashed is the FEG stopping,  orange-dotted curve is the SLPA-LM values~\cite{Peralta2022} for the inner-shells, dashed-double-dot curve, TDDFT results by Quashie \textit{et al.}~\cite{Quashie2018}.
Symbols: as in Fig.~\ref{hniv}, {$\Circle$}  medium-high energy experimental data in~\cite{iaea}.
}
\end{figure}

\begin{figure}[h]
\includegraphics[width=0.45\textwidth
]{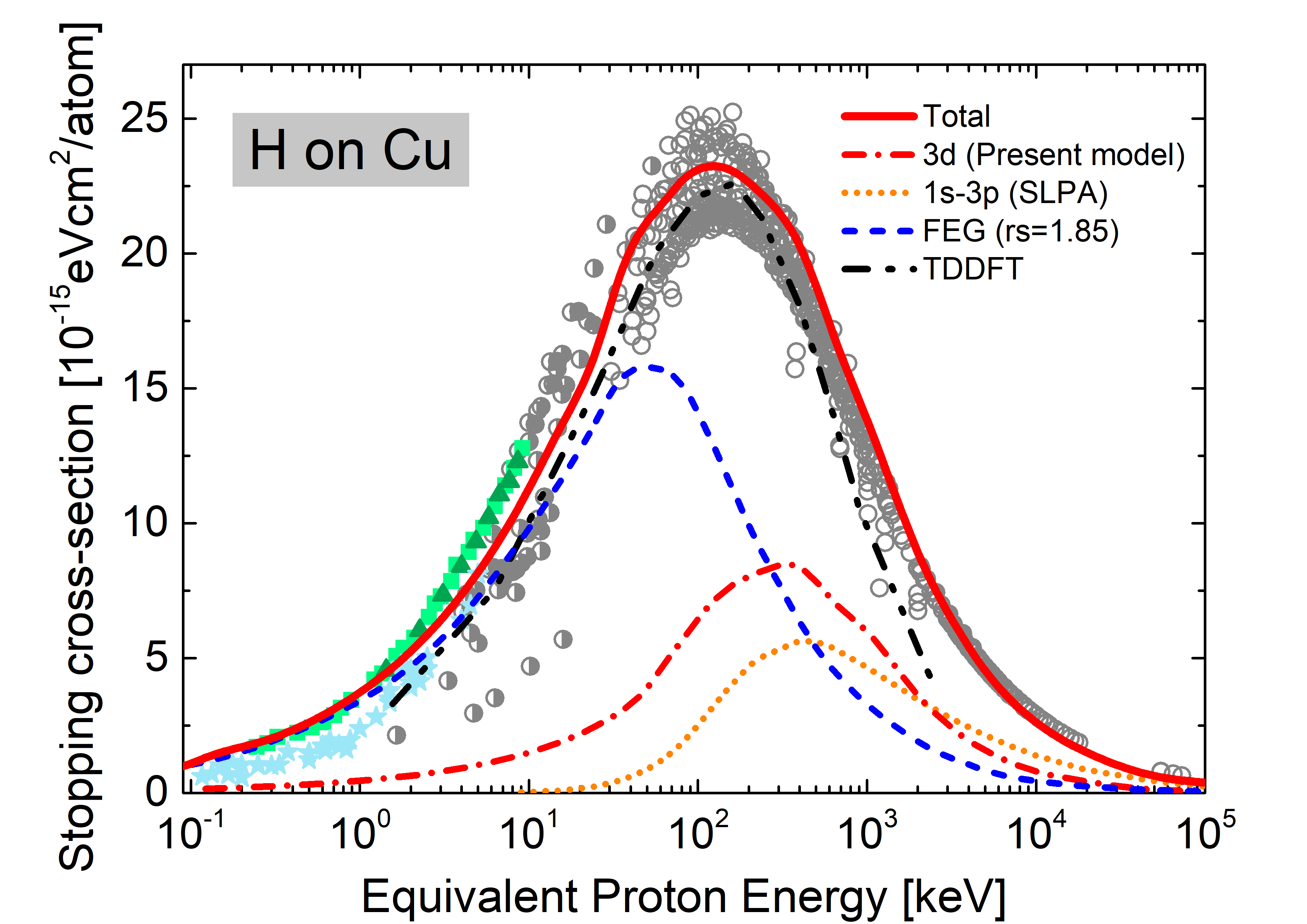}
\caption{\label{hcu_E} 
(Color online) Stopping cross-section of Cu for H as a function of the impact energy. 
Curves: as in Fig.~\ref{hni_E}, TDDFT results by Quashie \textit{et al.}~\cite{Quashie2016}.
Symbols: as in Fig.~\ref{hcuv}, 
{$\Circle$} experimental data in \cite{iaea}.
}
\end{figure}

\begin{figure}[h]
\includegraphics[width=0.45\textwidth
]{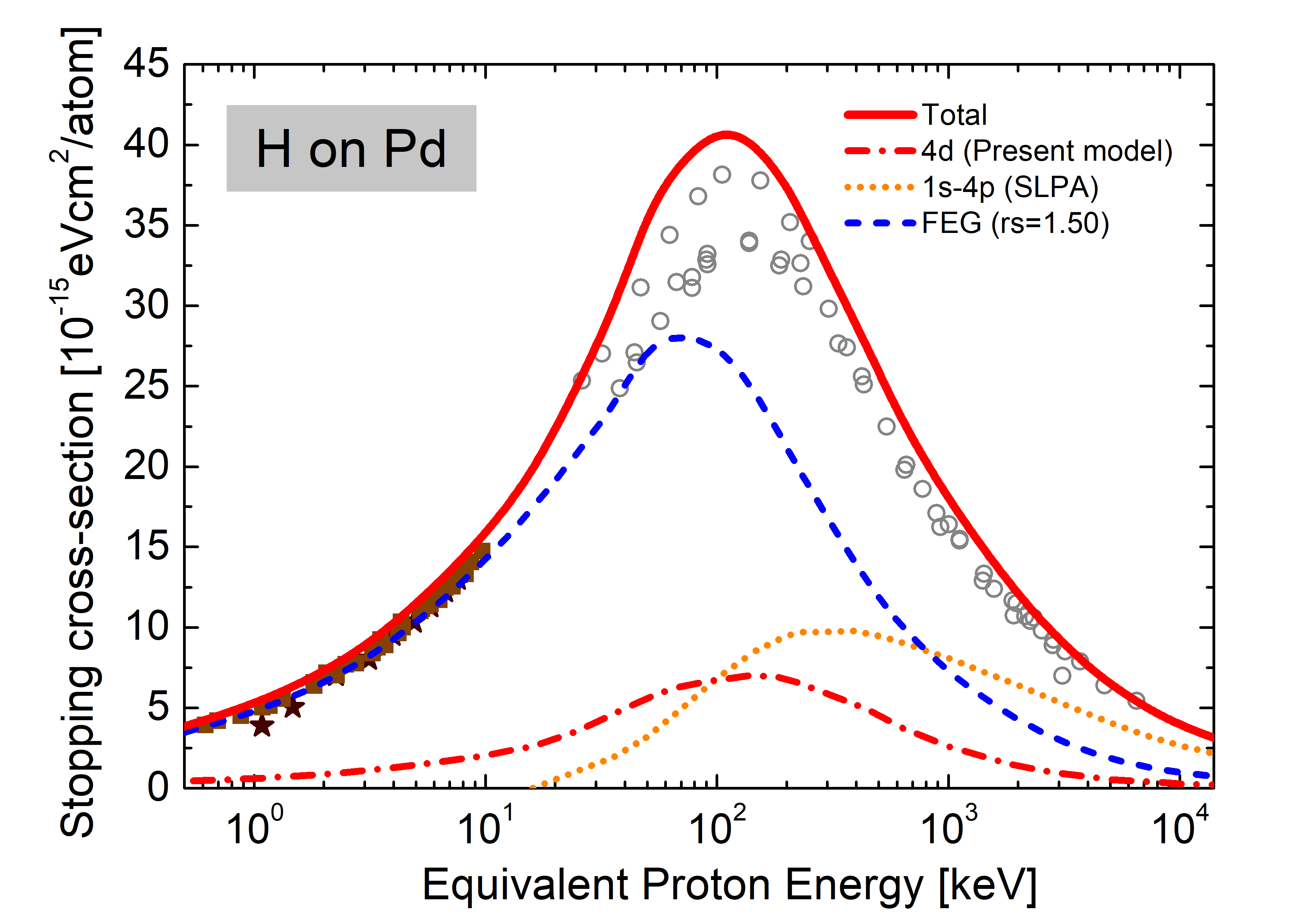}
\caption{\label{hpd_E} 
(Color online) Stopping cross-section of Pd for H as a function of the impact energy. Curves: as in Fig.~\ref{hni_E}.
Symbols: as in Fig.~\ref{hpdv},   
dark gray-solid curve is the SLPA-LM values~\cite{Peralta2022} for the inner-shells.
Symbols: as in Fig.~\ref{hpdv}, 
{$\Circle$} 
experimental data in \cite{iaea}.
}
\end{figure}

\begin{figure}[H]
\includegraphics[width=0.45\textwidth]{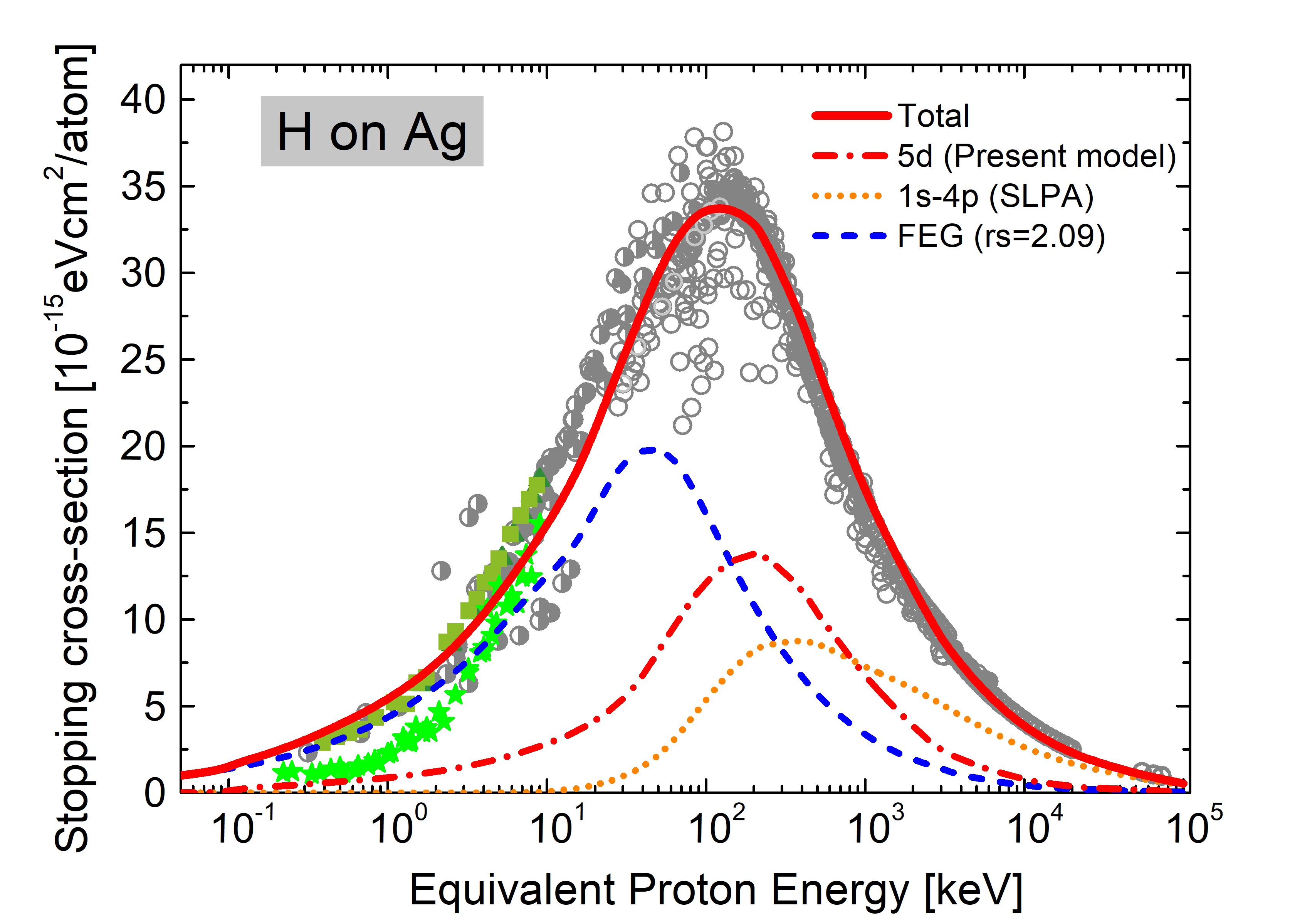}
\caption{\label{hag_E} 
(Color online) Stopping cross-section of Ag for H as a function of the impact energy. 
Curves: as in Fig.~\ref{hpd_E}.
Symbols: as in Fig.~\ref{hagv}, 
{$\Circle$} experimental data in \cite{iaea}.
}
\end{figure}

\begin{figure}
\includegraphics[width=0.45\textwidth
]{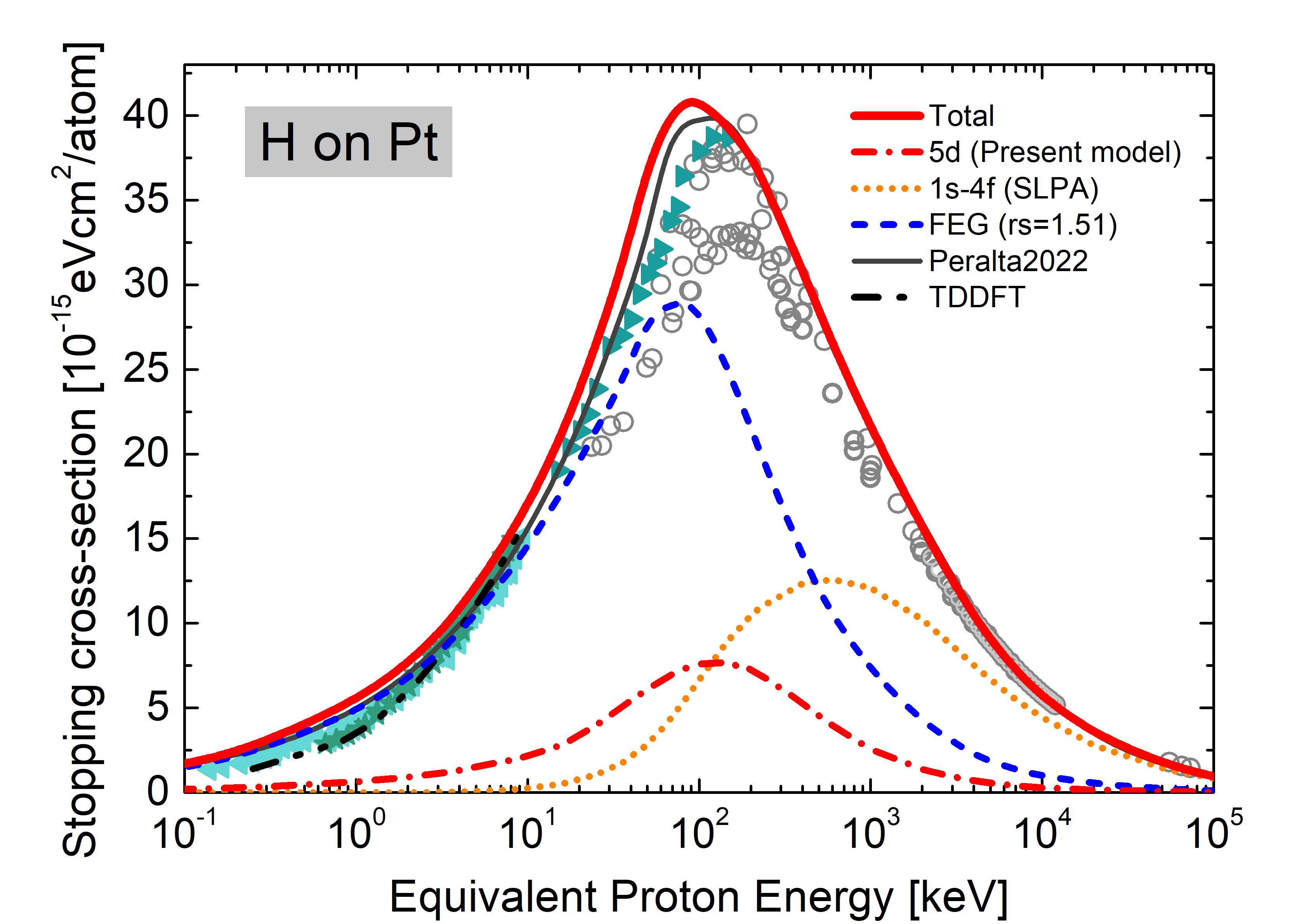}
\caption{\label{hpt_E} 
(Color online) Stopping cross-section of Pt for H as a function of the impact energy. 
Curves: as in Fig. \ref{hni_E}, with TDDFT curve in \cite{Li2023}.
Symbols: as in Fig.~\ref{hptv}
{$\Circle$} experimental data in \cite{iaea}.
}
\end{figure}

\begin{figure}[H]
\includegraphics[width=0.45\textwidth
]{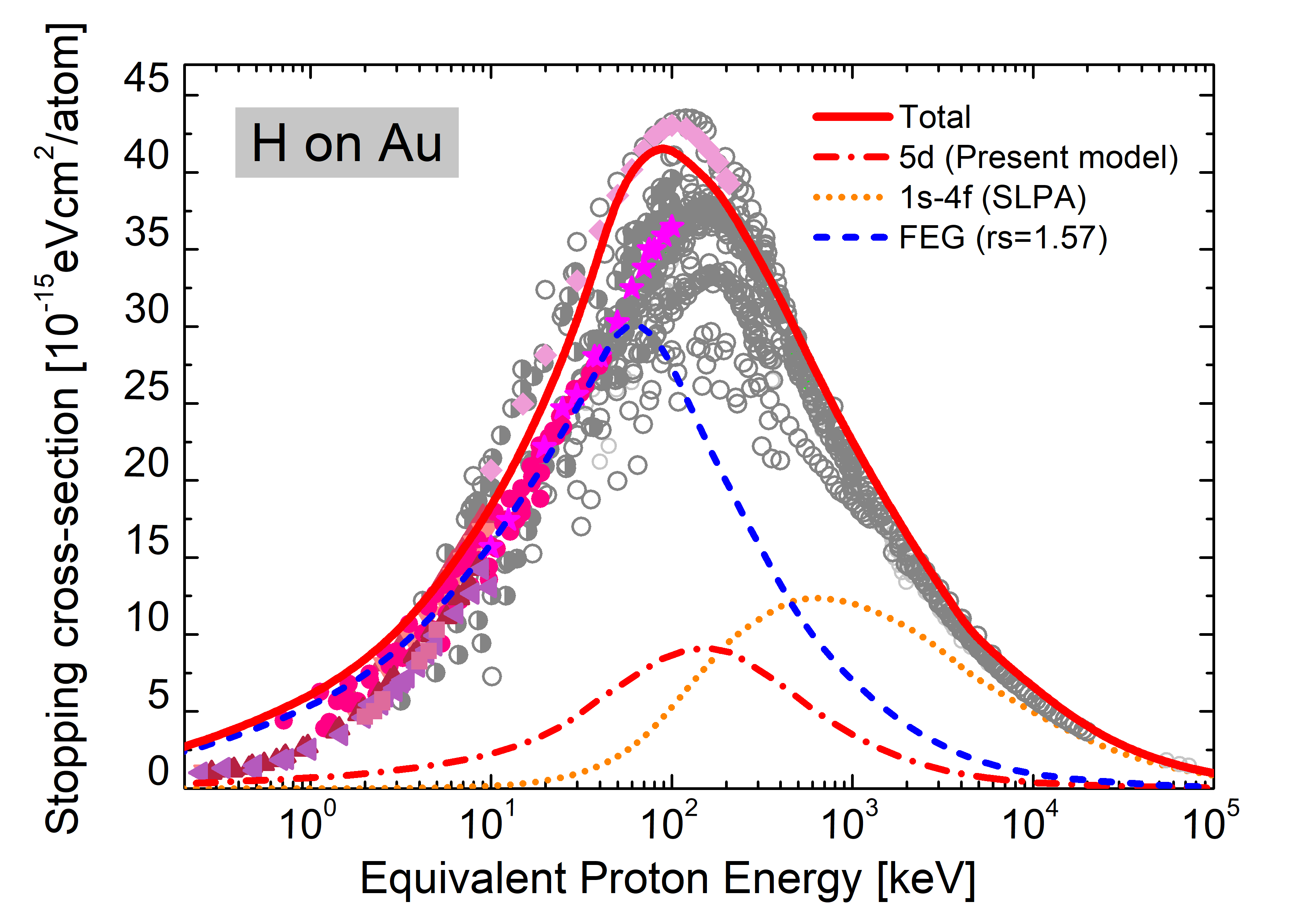}
\caption{\label{hau_E} 
(Color online) Stopping cross-section of Au for H as a function of the impact energy. 
Curves: as in Fig.~\ref{hpd_E}.
Symbols: as in Fig~\ref{hauv}, 
{$\Circle$} experimental data in \cite{iaea}.
}
\end{figure}

The energy loss in an extended energy range is displayed in Figs. \ref{hni_E} to \ref{hau_E}. In these figures, we present total and fully theoretical calculations that also include the SLPA-LM~\cite{Peralta2022} values for inner-shell contributions and the Mermin-Lindhard~\cite{mermin} results for the FEG for energies above that of plasmon excitation. Then, the present approach provides a coherent theoretical method capable of describing the stopping power from the very low ($0.1$~keV) to the high but not relativistic energy region ($100$~MeV). The whole picture given by Figs. \ref{hni_E} to \ref{hau_E} is promising. As already mentioned, the spread of experimental data around the stopping maximum characterizes the six targets, with Au, Ag, Cu, and Ni being the most striking cases.

For H in Ni, we display in Fig. \ref{hni_E} the extended in-energy version of the present calculations. In this figure, the contribution of the $3d$ electrons and of the $1s$ to $3p$ ones is shown in the extended energy range. It can be noted that the $3d$-curve has a maximum at $350$~keV/amu, 
where the ion velocity is close to the mean velocity of the $3d$-electrons, i.e. $v \simeq v_d$.
Although there is a striking disparity of values, the stopping maximum of H in Ni has not been measured since 1986  \cite{SE1986}. Our total stopping cross-section agrees very well with Ref.~\cite{Moller2002} and with the data around the stopping maximum from Ref. \cite{SE1986}.  The comparison with the TDDFT results by Quashie \textit{et al.}~\cite{Quashie2018} (off-channeling and 16 active electrons) is very good. It is worth noting that the present model includes all target electrons (up to the K-shell) by means of the SLPA-LM~\cite{Peralta2022}, allowing the description in the extended energy range shown in this figure.

The stopping cross sections for H in Cu as a function of the impact energy are displayed in Fig. \ref{hcu_E}. As mentioned for the case of Ni, the $3d$-contibution is maximum for $v \simeq v_d$. The present total stopping values are higher but in relatively good agreement with TDDFT results reported in Ref.~\cite{Quashie2016}. For energies above $500$~keV, the TDDFT curve underestimates the measurements due to the partial number of electrons considered. Again, the present agreement with the high-energy data is due to the inclusion of the $1s$ to $3p$ electrons contribution by using the SLPA-LM.

The results for H in Pd are presented in Fig. \ref{hpd_E}. This case is different from the previous ones: for impact energies above $100$~keV, the stopping power due to the $1s$ to $4p$ subshells is larger than the $4d$-subshell contribution. The $4d$-electrons of Pd have smaller mean velocities $v_d$ than Ni and Cu (see Table~\ref{table_period}). This is consistent with a maximum of the $d$-curve at lower energies ($E\sim 160$~keV). 
Our total values agree very well with the data at low and high energies, with the stopping maximum being a little overestimated ($\sim 7\%$ above the higher values).
Very recent measurements by researchers from the Uppsala University and Johannes Kepler University~\cite{Mor20} obtain the stopping maximum of H in Pd at $34\times 10^{-15}~$eVcm$^2$/atom, which is $15\%$ below our result. 
\vspace{-0.5cm} 
In Fig. \ref{hag_E}, we display the present stopping values of H in Ag. The agreement with the experimental data is good, considering the large dispersion of values. The present description of the $d$-electron contribution is crucial to the reported total stopping curve. The stopping power of H in Ag has more than forty different sets of measurements. However, the maximum is not experimentally well-defined, as shown in Fig. \ref{hag_E}. The latest measurements around the maximum were reported by Semrad and collaborators \cite{SE1986,Ep92} more than 30 years ago. It is worth mentioning that the present total stopping values agree very well with this group of measurements.

In Figs. \ref{hpt_E} and \ref{hau_E}, we display the present results for the stopping power of Pt and Au for H as a function of the impact energy, from $0.1$~keV to $100$~MeV. For Pt in Fig.~\ref{hpt_E}, our total stopping agrees well with the data at low and high energies. The stopping maximum is very sensitive in both theoretical models and experimental setups.
In this case, the present results around the maximum are close to the recent data from Refs.~\cite{Pr2012,Mor20,Sel20} with a slight overestimation on our model. We also include here our previous results~\cite{Peralta2022}, which includes 10 valence electrons in the FEG. We consider the present proposal of an \textit{inhomogeneous} distribution function for $d$-electrons to be more physically sound than the previous one. As in all the targets studied here, the maximum of the $d$ curve corresponds to $v\simeq v_d$. Similarly to Pd, Pt has a small mean velocity of the $d$-subshell. Concomitantly, the $d$-curve in Fig.~\ref{hpt_E} has a maximum at $E\sim 100$~keV, and for higher impact energies, the curve is lower than the contribution of the deepest $1s$ to $4f$ subshell. It is worth mentioning that the SLPA-LM results for these deep shells support the excellent agreement at high energies.

The case of H in Au is rather singular. This system has the largest number of stopping power measurements: 72 datasets and 1163 data points \cite{iaea,MONTANARI2024_IAEA}. This amount of values also implies having the largest experimental dispersion, as can be clearly noted in Fig.~\ref{hau_E}. As in the other cases, the present results have an overall good agreement with the experimental values in a very extensive energy range. It is fair to say that Au, with 79 electrons, is the most complex case considered here, and the response of all the 79 electrons has been taken into account. The agreement at high energies is very good, and the agreement at low and intermediate energies is good only with some datasets. The present total stopping overestimates the most recent data at low energies \cite{Markin2008} and around the maximum \cite{Pr2012}, being closer to previous measurements  \cite{MT1996}.

\section{\label{sec:concl} Conclusions}

In this work, we introduce a non-perturbative theoretical model to deal with the contribution of the $d$ electrons to the stopping power. The present proposal is based on the \textit{inhomogeneous}  momentum distribution function of the $d$-subshell. We found it especially suitable for the later transition metals whose $d$ electrons have very small binding energies. This feature allows a partial promotion to the conduction band, which causes the remaining bound electrons to contribute to the energy loss at very low impact velocities. 

We systematically studied the stopping power for the transition metals of groups 10 and 11: Ni, Cu, Pd, Ag, Pt, and Au. We examined the low energy dependence with the ion velocity of the energy loss and the total stopping power in an extended energy region. In the former, we obtained that the $d$-contribution is relevant at very low impact velocities. The sharp change of slope mentioned in experimental works is not found; instead, a soft non-linearity coherent with an inhomogeneous momentum distribution function is obtained. The extension to larger energies allows us to show that the present model has a maximum contribution at an impact velocity similar to the mean velocity of the $d$-electrons.

The agreement between our total stopping power and the experimental data at low energies is very good for group 10 transition metals. For the group 11 targets, the spread of data at low energies is large. Our results agree with some of the experimental values.  We do not rule out certain overestimations at very low impact velocities due to possible missing features, such as the energy gap of the remaining bound $d$ electrons. The comparison with available TDDFT at low energies is good and lets us validate and critically evaluate the present results.

We extend the description of the total stopping in a large energy region by including all the electronic contributions, from the FEG up to the deepest $1s$-electrons. To that end, we combine the present non-perturbative model for the $d$-subshell contribution with other approaches for the  FEG and for the inner shells. In this way, we provide a coherent theoretical method capable of describing the stopping power of the later transition metals from the very low to the high, yet not relativistic, energy region. The whole picture given by the six cases analyzed here is promising.

\begin{acknowledgments} 
We acknowledge Dr. Jorge Miraglia for his enlightened discussions about this work. The following institutions of Argentina financially support this research:
the Consejo Nacional de Investigaciones científicas y Técnicas (CONICET) by the project PIP11220200102421CO, the Agencia Nacional de Promoción Científica y Tecnológica (ANPCyT) by the project PICT-2020-SERIE A-01931.
\end{acknowledgments}

\bibliography{stopping}

\appendix
\section{Momentum distribution calculations}
\input{appendix6}\label{App}

\end{document}

%% file: appendix6.tex
%
%
%



In the present model, we propose an inhomogeneous momentum distribution function, $f(p)$, for the $d$-electrons given by 
\begin{equation}\label{eq:fn}
    f_{nl}(p)= \frac{(2 \pi)^3}{2}\ 
    |\Phi_{nl}(\vec{p})|^2, 
\end{equation}
with $\Phi_{nl}(p)$ being the Fourier transform of the radial wave function $\phi_{nl}(r)$, or $\phi_{nl\pm}(\vec{r})$ for the relativistic wave functions, with $nl$ being the quantum numbers, and in this case $l=2$.
We express the radial wave function in terms of Slater functions to have an analytical expression for  $\Phi_{nl}(p)$.

The radial wave function can be expanded in Slaters as
\begin{equation} \label{eq:fun}
    \phi_{n l}(r)
    =\sum_{j}  C_{j}\ R_{n_j l}(r) 
\end{equation} 
\noindent
with
\begin{equation} \label{eq:Sjl}
R_{n_j l}(r) = N_{j}\ r^{(n_{j}-1)} e^{-\zeta_{j}\ r},
\end{equation}
\noindent
and $N_{j}$ is given by
\begin{equation} \label{eq:Njl}
N_{j} = (2\zeta_{j})^{n_{j} + 1/2} / [(2n_{j})!]^{1/2}
\end{equation}
\noindent
The parameters $\zeta_{j}$, $C_{j}$ and $n_j$ are 
the coefficients of the Slater expansion and the main quantum number, respectively.

The Fourier transform of the wave function expressed in Eqs. (\ref{eq:fun}) to (\ref{eq:Njl}) is
\begin{equation}\label{phi1}
\begin{split}
\Phi_{nl}(\bar{p})
& = \frac{1}{(2 \pi)^{3/2}} \int  \phi_{n l}(r)\ e^{i \vec{p} \cdot \vec{r}} \ d \bar{r}  \\
& = \sum_j  A_{j} \int  r^{(n_{j}-1)} e^{-\zeta_{n_j l} r -i \vec{p} \cdot \vec{r}} 
\ d \bar{r}
\end{split}
\end{equation}
with $A_{j}=N_{j}\ C_{j}/(2 \pi)^{3/2}$.
We make use of the Flannery-Levy~\cite{Flannery} solution of this integral given by 
\begin{equation}\label{flann}
\begin{split}
     \int_0^{\infty} r^{\beta} & e^{-\alpha r+i \vec{p} \cdot \vec{r}} 
     \ d \bar{r} =  \\
 & 4 \pi\ i^l (\beta-l+1)! 
 \ F_l(p)
\end{split}
\end{equation}
\noindent
with 
\begin{equation}
    F_l(p)=\sum_{q=1}^{q_{\text {max }}} b_q \alpha^{s-l-q} \frac{p^l}{\left(\alpha^2+p^2\right)^s},
\end{equation}
$s =\beta + 3 - q $, $q_{\text {max }} = (\beta-l+2)/2$ or $(\beta-l+3)/2$ for $\beta-l$ being even or odd, and
$$
b_q  =(-1)^{q-1} \frac{(s-1)!}{(q-1)!} \frac{2^{s-q}}{(s-l-q)!}
$$

For the case of $d$ subshells studied here , 
by replacing \ref{flann} in \ref{phi1}, we obtain
\begin{equation}\label{ec:phi_nd}
    \Phi_{nd}(p)=
     \sum_{j}(-4 \pi) A_j (n-2)! 
     \ F_2(p).
\end{equation}

\noindent
Substituting Eq. (\ref{ec:phi_nd}) in Eq. (\ref{eq:trf}), and carrying out some algebraic calculations, the distribution function for the case of $3d$ electrons is

\begin{equation}\label{phi3}
\begin{split}
    f_{3d}(p)& =\frac{(2 \pi)^3}{2} |\Phi_{32}(p)|^2 \\
    & = \frac{(2 \pi)^3}{2} \frac{N_e}{4\pi} \frac{3^2 \ 2^9}{\pi} \sum_{i=1}^{5} \sum_{j=1}^{5} \frac{A_i A_j \zeta_j \zeta_j\ p^4}{(\zeta_i^2 + p^2)^4 (\zeta_j^2 + p^2)^4},
\end{split}
\end{equation}
where $N_e/4\pi$ comes from 
the normalization of the wave function to the number of $3d$ electrons.
Replacing (\ref{phi3}) in the integral $I(v_r)$ of the Eq. (\ref{eq:Ivr}) we obtain
\begin{equation}\label{I_A}
I(v_r) = \frac{(2 \pi)^3}{2} \frac{N_e}{4\pi} \frac{3^2 \ 2^9}{\pi} \sum_{i=1}^{5} \sum_{j=1}^{5} A_i A_j \zeta_i \zeta_j G(v_r)
\end{equation}
where
\begin{equation}
    G(v_r) = (v²_{r}+v^2)\ g_5(\zeta_i, \zeta_j, v_r) - g_7(\zeta_i, \zeta_j, v_r)
\end{equation}
with
\begin{equation}\label{g5}
   g_5(\zeta_i, \zeta_j, v_r) = \int_{|v_r - v|}^{{v}_r + v} \frac{p^5\ dp}{(\zeta_i^2 + p^2)^4 (\zeta_j^2 + p^2)^4} 
\end{equation}
and 
\begin{equation}\label{g7}
   g_7(\zeta_i, \zeta_j, p) = \int_{|v_r - v|}^{{v}_r + v} \frac{p^7 \ dp}{(\zeta_i^2 + p^2)^4 (\zeta_j^2 + p^2)^4} 
\end{equation}

The integrals in Eqs.~(\ref{g5}) and (\ref{g7}) are analytical, so Eq.~(\ref{eq:Ivr}) can be included in Eq.~(\ref{eq:nagy4_new}). Numerical integration is also possible, and in fact, we have checked both.

Analogously, the distributions $ f_{4d}(p) $ and $ f_{5d}(p) $ were obtained. Algebra follows the same steps, although the calculation is heavier in these cases due to the number of terms to consider. 
The expression of $f_{3d}(p)$ given by Eq.~(\ref{phi3}), can be generalized for all and obtain $f_{nd}(p)$ as linear combinations of functions $h(a,b, i, j)$ of the type
\begin{equation}
    h(a,b,i,j,p)=\frac{p^4}{(a^2+p^2)^i (b^2+p^2)^j},
\end{equation}
with $i=j=4$ for the $3d$ subshell, $i=4-5$ and $j=4-5$ for the $4d$, and $i=4-6$ and $j=4-6$ for the $5d$ subshell.

The integrals $I(v_r)$ of the Eq. (\ref{eq:Ivr})  could be calculated analytically, but in the two latter cases, it was done numerically.